\input harvmac
\input epsf
\def\ra{\rightarrow}
\def\epsK{{\varepsilon}_K}
\def\barB{{\overline B}}
\def\gsim{{~\raise.15em\hbox{$>$}\kern-.85em
          \lower.35em\hbox{$\sim$}~}}
\def\lsim{{~\raise.15em\hbox{$<$}\kern-.85em
          \lower.35em\hbox{$\sim$}~}} 
\def\Re{{\cal R}e}
\def\Im{{\cal I}m}

\def\YGTitle#1#2{\nopagenumbers\abstractfont\hsize=\hstitle\rightline{#1}%
\vskip .4in\centerline{\titlefont #2}\abstractfont\vskip .3in\pageno=0}
\YGTitle{MZ-TH/99-19, WIS-99/21/May-DPP, hep-ph/9905397}
{\vbox{
\centerline{Constraining New Physics with the}
\centerline{CDF Measurement of CP Violation in $B\ra\psi K_S$}}}
\bigskip
\centerline{Gabriela Barenboim}
\smallskip
\centerline{\it 
Institut f\"ur Physik - Theoretische Elementarteilchenphysik}
\centerline{\it Johannes Gutenberg-Universit\"at, D-55099 Mainz, Germany}
\centerline{gabriela@thep.physik.uni-mainz.de}
\bigskip
\centerline{Galit Eyal and Yosef Nir}
\smallskip
\centerline{\it Department of Particle Physics}
\centerline{\it Weizmann Institute of Science, Rehovot 76100, Israel}
\centerline{galit@wicc.weizmann.ac.il, ftnir@wicc.weizmann.ac.il}
\bigskip

\baselineskip 18pt
\noindent
Recently, the CDF collaboration has reported a measurement of
the CP asymmetry in the $B\rightarrow\psi K_S$ decay:
$a_{\psi K_S}=0.79^{+0.41}_{-0.44}$. We analyze the constraints
that follow from this measurement on the size and the phase
of contributions from new physics to $B-\barB$ mixing.
Defining the relative phase between the full $M_{12}$ amplitude
and the Standard Model contribution to be $2\theta_d$, we
find a new bound: $\sin2\theta_d\gsim-0.6\ (-0.87)$ at one sigma
(95\% CL). Further implications
for the CP asymmetry in semileptonic $B$ decays are discussed.

\bigskip
 
\baselineskip 18pt
\leftskip=0cm\rightskip=0cm
 
\Date{}


Recently, the CDF collaboration has reported a 
measurement of the CP asymmetry in the $B\ra\psi K_S$ decay
\ref\CDF{T. Affolder {\it et al.}, CDF collaboration, hep-ex/9909003.}:
\eqn\CDFnum{a_{\psi K_S}=0.79^{+0.41}_{-0.44},}
where
\eqn\defapk{{\Gamma(\barB^0_{\rm phys}(t)\ra\psi K_S)-
\Gamma(B^0_{\rm phys}(t)\ra\psi K_S)\over
\Gamma(\barB^0_{\rm phys}(t)\ra\psi K_S)+
\Gamma(B^0_{\rm phys}(t)\ra\psi K_S)}=a_{\psi K_S}\sin(\Delta m_Bt).}
(Previous searches have been reported by OPAL
\ref\OPALpre{K. Ackerstaff {\it et al.},
 Eur. Phys. J. C5 (1998) 379, hep-ex/9801022.}\
and by CDF
\ref\CDFpre{F. Abe {\it et al.},
 Phys. Rev. Lett. 81 (1998) 5513, hep-ex/9806025.}.)
Within the Standard Model, the value of $a_{\psi K_S}$ can be cleanly
interpreted in terms of the angle $\beta$ of the unitarity triangle,
$a_{\psi K_S}=\sin2\beta$. The resulting constraint is still weak,
however, compared to the indirect bounds from measurements of 
$|V_{ub}/V_{cb}|$, $\Delta m_B$ and $\epsK$
\ref\BPB{The BaBar Physics Book, eds. P.F. Harrison and H.R. Quinn,
SLAC-R-504 (1998).}:
\eqn\BABcon{\sin2\beta\in[+0.4,+0.8].}
Yet, the CDF measurement is quite powerful in constraining contributions
from new physics to the $B-\bar B$ mixing amplitude. It is the purpose
of this work to investigate this constraint. 

We focus our analysis on a large class of models of new physics
with the following features:
\item{(i)} The $3\times3$ CKM matrix is unitary. In particular,
the following unitarity relation is satisfied:
\eqn\Unidb{V_{ud}V_{ub}^*+V_{cd}V_{cb}^*+V_{td}V_{tb}^*=0.}
\item{(ii)} Tree-level decays are dominated by the Standard Model
contributions. In particular, the phase of the $\bar B\ra\psi K_S$ decay
amplitude is given by the Standard Model CKM phase, $\arg(V_{cb}V_{cs}^*)$,
and the following bound, which is based on measurements of Standard Model 
tree level processes only, is satisfied:
\eqn\uoverc{R_u\equiv
\left|{V_{ud}V_{ub}^*\over V_{cd}V_{cb}^*}\right|\lsim0.45.}

\noindent The first assumption is satisfied by all models with only three
quark generations (that is, neither fourth generation quarks
nor quarks in vector-like representations of the Standard Model).
The second assumption is satisfied in many extensions of the
Standard Model, such as most models of supersymmetry with $R$-parity and
left-right symmetric (LRS) models. There exist, however, viable models
where this assumption may fail, such as supersymmetry without $R$-parity
(see, for example, the discussion in
\nref\GrWo{Y. Grossman and M.P. Worah,
 Phys. Lett. B395 (1997) 241, hep-ph/9612269.}%
\nref\BaSt{R. Barbieri and A. Strumia,
 Nucl. Phys. B508 (1997) 3, hep-ph/9704402.}%
\refs{\GrWo,\BaSt}
or specific multi-scalar models 
\ref\KSW{K. Kiers, A. Soni and G.-H. Wu, Phys. Rev. D59 (1999) 096001,
 hep-ph/9810552.}).
Within the class of models that satisfies (i) and (ii), 
our analysis is model-independent.

The effect of new physics that we are interested in is the
contribution to the $B-\barB$ mixing amplitude,
$M_{12}-{i\over2}\Gamma_{12}$. Our second assumption implies that 
\eqn\NPGamma{\Gamma_{12}\approx\Gamma_{12}^{\rm SM}.}
The modification of $M_{12}$ can be parameterized as follows
\nref\GNW{Y. Grossman, Y. Nir and M.P. Worah,
 Phys. Lett. B407 (1997) 307, hep-ph/9704287.}%
(see, for example, \refs{\GNW,\BPB}):
\eqn\defrthe{M_{12}=r_d^2 e^{2i\theta_d}M_{12}^{\rm SM}.}
The experimental measurement of $\Delta m_B$ provides bounds
on $r_d^2$ while the new CDF measurement of $a_{\psi K_S}$
gives the first constraint on $2\theta_d$.

The implications for CP violation in $B$ decays of models with
the above features has been discussed in refs. 
\nref\NiSi{Y. Nir and D. Silverman, Nucl. Phys. B345 (1990) 301.}%
\nref\DLN{C.O. Dib, D. London and Y. Nir,
 Int. J. Mod. Phys. A6 (1991) 1253.}%
\nref\SoWo{J.M. Soares and L. Wolfenstein, Phys. Rev. D47 (1993) 1021.}%
\nref\Nixs{Y. Nir, Phys. Lett. B327 (1994) 85, hep-ph/9402348.}%
\nref\Gros{Y. Grossman, Phys. Lett. B380 (1996) 99, hep-ph/9603244.}%
\nref\DDO{N.G. Deshpande, B. Dutta and S. Oh,
 Phys. Rev. Lett. 77 (1996) 4499, hep-ph/9608231.}%
\nref\SiWo{J.P. Silva and L. Wolfenstein,
 Phys. Rev. D55 (1997) 5331, hep-ph/9610208.}%
\nref\CKLN{A.G. Cohen, D.B. Kaplan, F. Lepeintre and A.E. Nelson, 
 Phys. Rev. Lett. 78 (1997) 2300, hep-ph/9610252.}%
\nref\BBBV{G. Barenboim, F.J.Botella, G.C.Branco and O.Vives, 
 Phys. Lett. B422 (1998) 277, hep-ph/9709369.}%
\refs{\NiSi-\BBBV}. Analyses that are similar to ours have also
appeared, prior to the CDF measurement, in refs.  
\nref\GKOT{T. Goto, N. Kitazawa, Y. Okada and M. Tanaka,
 Phys. Rev. D53 (1996) 6662, hep-ph/9506311.}%
\nref\Xing{Z. Xing, Eur. Phys. J. C4 (1998) 283, hep-ph/9705358.}%
\nref\SaXi{A.I. Sanda and Z. Xing,
 Phys. Rev. D56 (1997) 6866, hep-ph/9708220.}%
\nref\LWol{L. Wolfenstein, Phys. Rev. D57 (1998) 6857, hep-ph/9801386.}%
\nref\RaSu{L. Randall and S. Su,
 Nucl. Phys. B540 (1999) 37, hep-ph/9807377.}%
\nref\CaWo{R.N. Cahn and M.P. Worah, hep-ph/9904480.}%
\refs{\GKOT-\CaWo,\GNW}. 

To derive bounds on $r_d^2$ and $2\theta_d$ we need to know
the allowed range for the relevant CKM parameters.
Assuming CKM unitarity \Unidb\ and Standard Model dominance in tree
decays \uoverc, we get:
\eqn\VtdSM{0.005\lsim|V_{td}V_{tb}^*|\lsim0.013,}
\eqn\betaSM{0\lsim\beta\lsim\pi/6\ \ {\rm or}\ \ 5\pi/6\lsim\beta\lsim2\pi.}
Note that these ranges are much larger than the Standard Model ranges.
The reason for that is that we do not use here the $\Delta m_B$
and $\epsK$ constraints. These are loop processes and, in our framework,
could receive large contributions from new physics.

Let us first update the constraint on $r_d^2$. To do so, we write
the Standard Model contribution to $\Delta m_B$ in the following way
\nref\BBL{G. Buchalla, A.J. Buras and M.E. Lautenbacher,
 Rev. Mod. Phys. 68 (1996) 1125, hep-ph/9512380.}%
(see \refs{\BBL,\BPB}\ for definitions and numerical values
of the relevant parameters):
\eqn\DMB{\left[{2M_{12}^{\rm SM}\over 0.471\ ps^{-1}}\right]=
\left[{\eta_B\over0.55}\right]
\left[{S_0(x_t)\over2.36}\right]
\left[{f_{B_d}\sqrt{B_{B_d}}\over0.2\ GeV}\right]^2
\left[{V_{td}V_{tb}^*\over8.6\times10^{-3}}\right]^2.}
The main uncertainties in this calculation come from
eq. \VtdSM\ and from
\eqn\BBfB{f_{B_d}\sqrt{B_{B_d}}\ =\ 160-240\ MeV.}
Using
\eqn\DMBfull{\Delta m_B=r_d^2|2M_{12}^{\rm SM}|,}
we find:
\eqn\fullSM{0.3\lsim r_d^2\lsim5.}

Next we derive the new constraint on $2\theta_d$. With the
parameterization \defrthe, we have
\eqn\apkSM{a_{\psi K_S}=\sin2(\beta+\theta_d).}
Defining
\eqn\defma{\eqalign{
\beta_{\rm max}\ \equiv&\ \arcsin[(R_u)_{\rm max}],\cr
2\bar\beta_{\min}\ \equiv&\ \arcsin[(a_{\psi K_S})_{\rm min}],\cr}}
where both $\beta_{\rm max}$ and $\bar\beta_{\min}$ are defined to lie
in the first quadrant, we find that the following range for $2\theta_d$
is allowed:
\eqn\excthe{2(\bar\beta_{\min}-\beta_{\rm max})\leq2\theta_d\leq
\pi+2(\beta_{\rm max}-\bar\beta_{\min}).}
The constraint \excthe\ can be written simply as
\eqn\sinthe{\sin2\theta_d\geq-\sin2(\beta_{\rm max}-\bar\beta_{\min}).}

Within our framework, the allowed range for $\beta$ is given in \betaSM,
that is $2\beta_{\rm max}\approx\pi/3$.
Taking the CDF measurement \CDFnum\ to imply, at the one sigma level,
\eqn\mild{a_{\psi K_S}\gsim0.35,}
or, equivalently,
\eqn\bethco{2\bar\beta_{\rm min}\approx\pi/9,}
we find $2(\beta_{\rm max}-\bar\beta_{\min})\approx2\pi/9$
and, consequently, 
\eqn\mildbo{\sin2\theta_d\gsim-0.6.}
If we take a more conservative approach and consider the 95\%  CL
lower bound,
\eqn\stro{a_{\psi K_S}\geq0,}
or, equivalently,
\eqn\stroco{2\bar\beta_{\rm min}\approx0,}
we find $2(\beta_{\rm max}-\bar\beta_{\min})\approx\pi/3$
and, consequently, 
\eqn\strobo{\sin2\theta_d\gsim-0.87.}
Eq. \mildbo\ (or the milder constraint \strobo), being the first
constraint on $\theta_d$, is our main result.
 
There are two main ingredients in the derivation of the bounds \mildbo\
and \strobo. The validity of one of them, that is the bound on 
$\sin2(\beta+\theta_d)$ from the value of $a_{\psi K_S}$, depends on the 
size of contributions to the $b\ra c\bar cs$ decay that carry a phase that 
is different from $\arg(V_{cb}V_{cs}^*)$. To understand the effects of such 
new contributions, we define
\eqn\nptree{\theta_A=\arg(\bar A_{\psi K_S}/\bar A^{\rm SM}_{\psi K_S}),}
where $\bar A_{\psi K_S}$ is the $\bar B\ra\psi K_S$ decay amplitude.
For $\theta_A\neq0$, eq. \apkSM\ is modified into
\eqn\apktree{a_{\psi K_S}=\sin2(\beta+\theta_d+\theta_A).}
The bounds \mildbo\ and \strobo\ apply now to the combination of new phases
$\theta_d+\theta_A$. Since, however, $|\sin\theta_A|\leq
|\bar A^{\rm NP}_{\psi K_S}/\bar A^{\rm SM}_{\psi K_S}|$,
we expect $\theta_A$ to be small. Then, we can still 
use \mildbo\ and \strobo, with the right hand side relaxed by
${\cal O}(\theta_A)$, as lower bounds on $\sin2\theta_d$.
Examining the actual numerical values of the bounds \mildbo\ and \strobo,
we learn that for $|\bar A^{\rm NP}_{\psi K_S}/\bar A^{\rm SM}_{\psi K_S}|
\lsim0.01$, the effect is clearly unimportant. It takes a very large new 
contribution, $|\bar A^{\rm NP}_{\psi K_S}/\bar A^{\rm SM}_{\psi K_S}|\gsim
0.4(0.25)$, to completely wash away our one sigma (95\% CL)
bounds. We are not familiar with any reasonable extension of the
Standard Model where the new contribution is that large.
For example, in the framework of supersymmetry with $R_p$,
a model independent analysis of supersymmetric contributions
to the $b\ra c\bar cs$ decay
\ref\CFMMS{M. Ciuchini, E. Franco, G. Martinelli, A. Masiero and
 L. Silvestrini,  Phys. Rev. Lett. 79 (1997) 978, hep-ph/9704274.}\
finds an upper bound, 
$|\bar A^{\rm SUSY}_{\psi K_S}/\bar A^{\rm SM}_{\psi K_S}|\lsim0.1$. 
The bound can be saturated only with light supersymmetric spectrum and 
maximal flavor changing gluino couplings. In most supersymmetric
flavor models, however, the relevant coupling is of order $|V_{cb}|$
and $|\bar A^{\rm SUSY}_{\psi K_S}/\bar A^{\rm SM}_{\psi K_S}|$ is well below 
the percent level. This is the case, for example, in models of universal 
squark masses, of alignment and of non-Abelian horizontal symmetries 
(see {\it e.g.} ref. \BaSt). In LRS models, with $m(W_R)\gsim1\ TeV$ and 
$|V^R_{cb}|\sim|V^L_{cb}|$, we have $|\bar A^{\rm LRS}_{\psi K_S}/
\bar A^{\rm SM}_{\psi K_S}|\lsim0.01$. 

The other ingredient of our analysis, 
that is the bound on $\sin\beta$ from $R_u$, suffers from hadronic 
uncertanties in the determination of the allowed range for
$R_u$. We have used $|V_{ub}/V_{cb}|\lsim0.10$. We emphasize, however,
that uncontrolled theoretical errors, that is the hadronic modelling
of charmless $B$ decays, are the main source of uncertainty in
determining the range for $|V_{ub}/V_{cb}|$. It would
be misleading then to assign a confidence level to our bound on
$\sin2\theta_d$. (See a detailed discussion in ref. \BPB.)
All we can say is that if indeed $|V_{ub}/V_{cb}|\leq0.10$ holds,
as suggested by various hadronic models, then $\sin2\theta_d\geq-0.6(-0.87)$
at one sigma  (95\% CL). The measurement of $a_{\psi K_S}$ would not
provide any bound on $\sin2\theta_d$ at one sigma (95\% CL) if
$|V_{ub}/V_{cb}|$ were as large as 0.17 (0.15).

When investigating specific models of new physics, it is often
convenient to use a different parameterization of the new
contributions to $M_{12}$. Instead of \defrthe, one uses
(see, for example, \RaSu\ in the supersymmetric framework and
\ref\Bare{G. Barenboim, Phys. Lett. B443 (1998) 317, hep-ph/9810325.}\ 
in the left-right symmetric framework):
\eqn\defrthe{M_{12}^{\rm NP}=h e^{i\sigma}M_{12}^{\rm SM},}
where $M_{12}^{\rm NP}$ is the new physics contribution.
The relation between the two parametrizations is given by
\eqn\twopar{r_d^2 e^{2i\theta_d}=1+h e^{i\sigma}.}
To derive the CDF constraints in the $(h,\sigma)$ plane, the
following relations are useful:
\eqn\usefu{r_d^2=\sqrt{1+2h\cos\sigma+h^2}.}
\eqn\useful{\sin2\theta_d={h\sin\sigma\over
\sqrt{1+2h\cos\sigma+h^2}}.}

The bound of eq. \fullSM\ corresponds to the allowed region
in the ($h,\sigma$) plane presented in Figure 1. 
\medskip

\centerline{\epsfbox{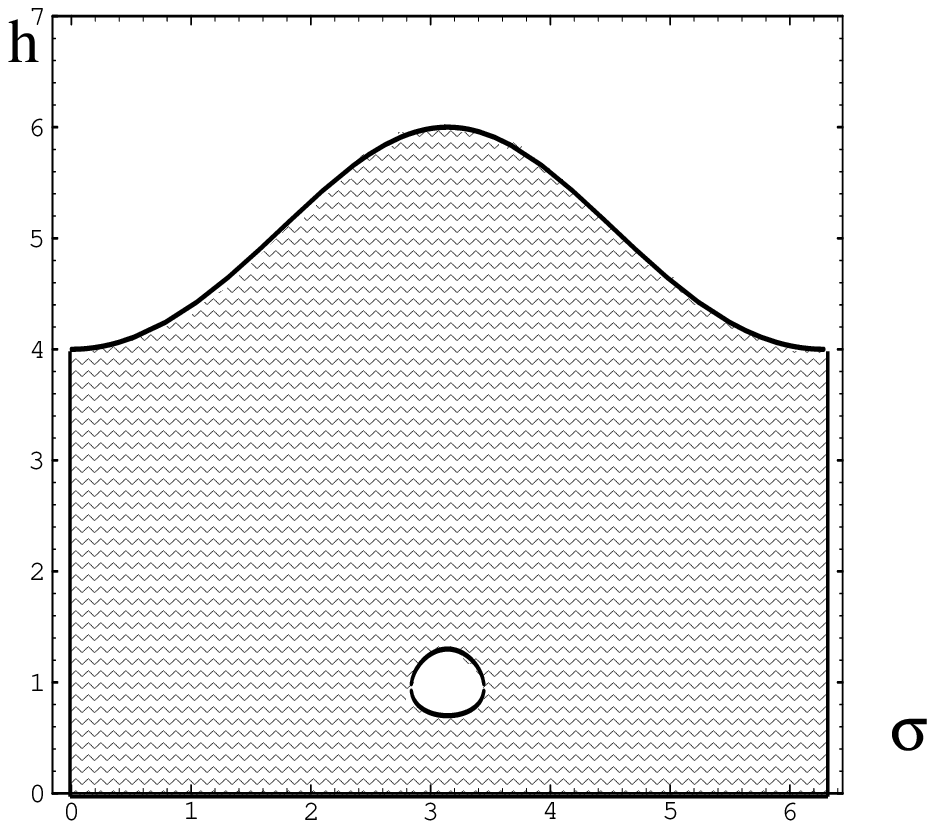}}

{\rm Figure 1. } The $\Delta m_B$ constraint. The grey region is allowed. 
\medskip


The situation is particularly
interesting for values of $\sigma$ close to $\pi$. Here,
the Standard Model and the new physics contributions add
destructively. Consequently, large values of $h$ up to
\eqn\hmax{h_{\rm max}=(r_d^2)_{max}+1\approx6}
are allowed; this means that new physics may still be
dominant in $B-\barB$ mixing. On the other hand, values
of $h$ close to 1 are forbidden since the new physics
contribution cancels the Standard Model amplitude, yielding
values of $\Delta m_B$ that are too small. 

The bounds of eqs. \mildbo\ and \strobo\ are presented in Figure 2. 
\medskip

\centerline{\epsfbox{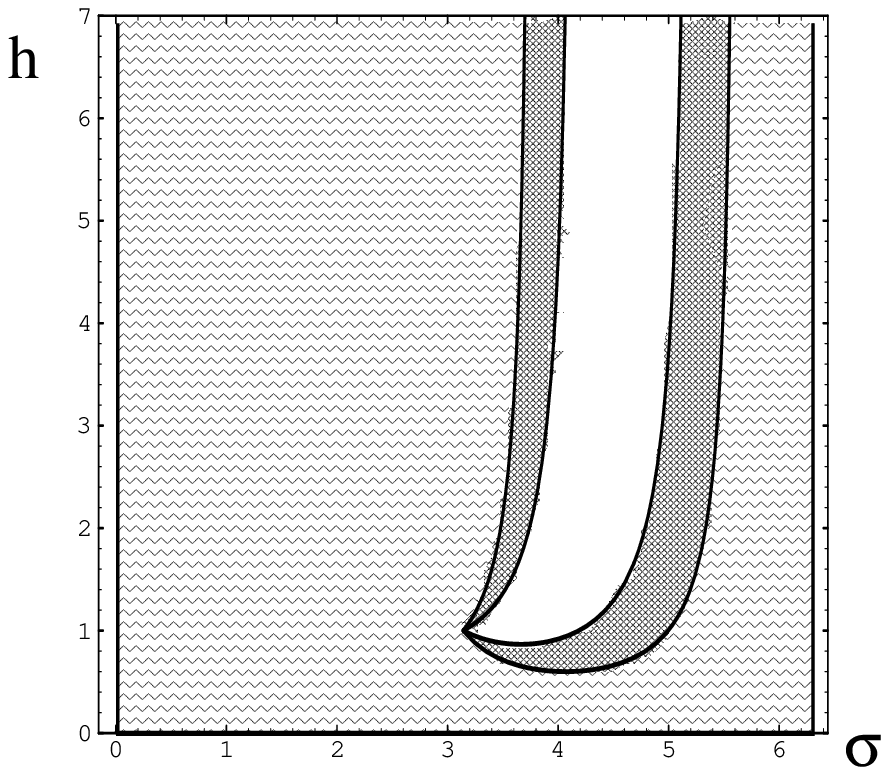}}

{\rm Figure 2. } The $a_{\psi K_S}$ constraint. The allowed region
corresponding to the one sigma (95\% CL) bound, $a_{\psi K_S}\geq0.35\
(0)$, is given by the light (light plus dark) grey area.
\medskip


We would like to emphasize some features of the excluded region:
\item{1.} Since only negative $\sin2\theta_d$ values are excluded,
only negative $\sin\sigma$ values are excluded.
\item{2.} For very large $h$, the Standard Model contribution
is negligible and, consequently, $\sin\sigma\approx\sin2\theta_d$.
Therefore, for large $h$ values, $\sigma$-values in the range
$[\pi+2(\beta_{\rm max}-\bar\beta_{\min}), 
2\pi-2(\beta_{\rm max}-\bar\beta_{\min})]$ are excluded.
\item{3.} For $\sigma$ arbitrarily close to $\pi$ (from above),
there is always an excluded region corresponding to $h$ similarly
close to 1. 

Finally, in Figure 3 we show the combination of the $\Delta m_B$
and $a_{\psi K_S}$ bounds. It is obvious that the latter adds
a significant exclusion region in the $(h,\sigma)$ plane.
\medskip

\centerline{\epsfbox{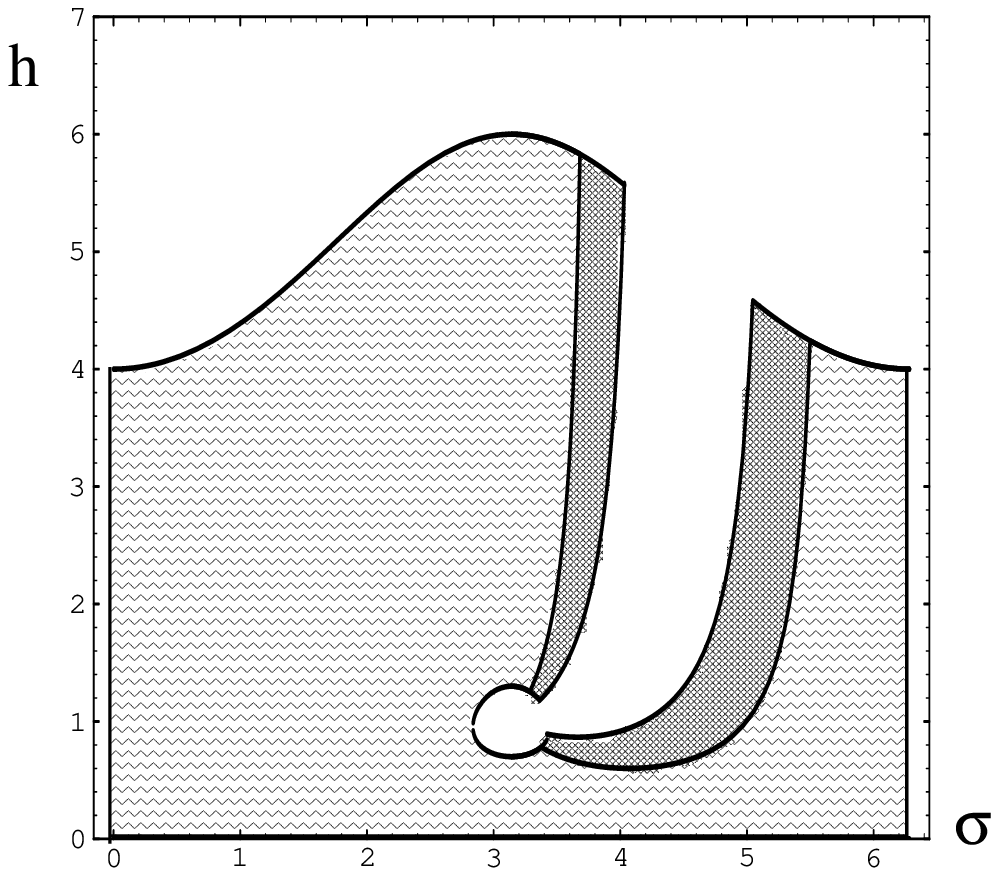}}

{\rm Figure 3. } The combination of the $\Delta m_B$ and $a_{\psi K_S}$
constraints. The light (light plus dark) grey region is the allowed region
corresponding to the one sigma (95\% CL) bound, $a_{\psi K_S}\geq0.35\
(0)$.
\medskip


The parameters that we have constrained here are related to other
physical observables. The ratio between the difference in decay
width and the mass difference between the two neutral $B$ mesons,
$\Delta\Gamma_B/\Delta m_B$, and the CP asymmetry in semileptonic 
decays, $a_{\rm SL}$, are given by
\eqn\DGDM{\eqalign{
{\Delta\Gamma_B\over\Delta m_B}\ =&\ \Re{\Gamma_{12}\over M_{12}},\cr
a_{\rm SL}\ =&\ \Im{\Gamma_{12}\over M_{12}}.\cr}}
The Standard Model value of $\Gamma_{12}/M_{12}$
has been estimated
\nref\Hage{J.S. Hagelin, Nucl. Phys. B193 (1981) 123.}%
\nref\BSS{A.J. Buras, W. Slominski and H. Steger, 
 Nucl. Phys. B245 (1984) 369.}%
\nref\BBD{M. Beneke, G. Buchalla and I. Dunietz,
 Phys. Rev. D54 (1996) 4419, hep-ph/9605259.}%
\refs{\Hage-\BBD,\CaWo}:
\eqn\SMGtoM{\left({\Gamma_{12}\over M_{12}}\right)^{\rm SM}\ \approx\ 
-(0.8\pm0.2)\times10^{-2},}
\eqn\argSM{\arg\left({\Gamma_{12}\over M_{12}}\right)^{\rm SM}\ =\ 
{\cal O}\left({m_c^2\over m_b^2}\right).}
We emphasize that there is a large hadronic uncertainty in this
estimate, related to the assumption of quark-hadron duality.
Eq. \SMGtoM\ leads to the following estimates: 
\eqn\SMpred{\eqalign{
\left|\left(\Delta\Gamma_B/\Delta m_B\right)^{\rm SM}\right|\ 
\sim&\ 10^{-2},\cr 
|(a_{\rm SL})^{\rm SM}|\ \lsim\ 10^{-3}.\cr}}
The (possible) measurement of $a_{\rm SL}$ can be used to
constrain the Standard Model CKM parameters \CaWo.

Since $(\Gamma_{12}/M_{12})^{\rm SM}$ is real to a good approximation,
the effects of new physics, within our framework, can be written as follows:
\eqn\aSLNP{\eqalign{
{\Delta\Gamma_B\over\Delta m_B}\ =&\ 
\left({\Gamma_{12}\over M_{12}}\right)^{\rm SM}{\cos2\theta_d\over r_d^2},\cr
a_{\rm SL}\ =&\ -\left({\Gamma_{12}\over M_{12}}\right)^{\rm SM}
{\sin2\theta_d\over r_d^2}.\cr}}
Note the following relation between the two observables:
\eqn\dGaSL{\sqrt{(\Delta\Gamma_B/\Delta m_B)^2+(a_{\rm SL})^2}=
\left|{\Gamma_{12}\over M_{12}}\right|^{\rm SM}{1\over r_d^2}.}
The lower bound on $r_d^2$ in eq. \fullSM\ implies then that
neither $\Delta\Gamma_B/\Delta m_B$ nor $a_{\rm SL}$ can be
enhanced compared to $(\Gamma_{12}/M_{12})^{\rm SM}$ by more
than a factor of about 3, that is a value of approximately
$3\times10^{-2}$. Moreover, if one of them is very
close to this upper bound, the other is suppressed.
(This is actually the situation within the Standard Model:
$\Delta\Gamma_B/\Delta m_B$ saturates the upper bound with $r_d=1$,
and $a_{\rm SL}$ is highly suppressed.) 

The new bound on $\sin2\theta_d$ that we found, eq. \mildbo\
(or the milder bound \strobo), do not affect the allowed region
for $\Delta\Gamma_B/\Delta m_B$. The reason is that $\cos2\theta_d$
is not constrained and could take any value in the range $[-1,+1]$.
On the other hand, the range for $a_{\rm SL}$ is affected.
Taking into account also the lower bound on $r_d^2$ in \fullSM, we find
\eqn\newaSL{
-3.3\lsim{a_{\rm SL}\over(\Gamma_{12}/M_{12})^{\rm SM}}\lsim2.0.}
The reduction in the upper bound from 3.3 to 2.0 is due to the
$a_{\psi K_S}$ bound. Note that $(\Gamma_{12}/M_{12})^{\rm SM}$ is
negative, so that the $a_{\psi K_S}$ constraint is a restriction on
negative $a_{\rm SL}$ values.

Similar analyses will be possible in the future for the $B_s$ system.
At present, there is only a lower bound on $\Delta m_{B_s}$,
\eqn\bouBs{\Delta m_{B_s}\ \geq\ 12.4\ {\rm ps}^{-1}.}
The main hadronic uncertainty comes from the matrix element,
\eqn\ranfBs{f_{B_s}\sqrt{B_{B_s}}\ =\ 200-280\ MeV.}
We find
\eqn\conrs{r_s^2\gsim0.6.}
Consequently, $|a_{\rm SL}(B_s)|$ is constrained to be smaller
than 1.6 times the Standard Model value of $|\Gamma_{12}(B_s)/M_{12}(B_s)|$.

Once an upper bound on a CP asymmetry in $B_s$ decay into a final
CP eigenstate is established, we will be able to constrain $2\theta_s$.
It will be particularly useful to use $b\rightarrow c\bar cs$ decays,
such as $B_s\ra D_s^+D_s^-$. The Standard Model value,
$a_{B_s\ra D_s^+D_s^-}\approx\sin2\beta_s$, is very small,
$\beta_s\equiv\arg[-(V_{ts}V_{tb}^*)/(V_{cs}V_{cb}^*)]={\cal O}(10^{-2})$.
Therefore, the Standard Model contribution can be neglected when
the bounds on $a_{B_s\ra D_s^+D_s^-}$ are well above the percent level.  
The approximate relation, $a_{B_s\ra D_s^+D_s^-}\approx-\sin2\theta_s$, 
will make the extraction of a constraint on $\sin2\theta_s$ particularly 
clean and powerful.

To summarize our main results: the CDF measurement of the CP asymmetry
in $B\rightarrow\psi K_S$ constrains the size and the phase of new physics
contributions to $B-\bar B$ mixing. The constraints are depicted in
Figures 2 and 3. They can be written as a lower bound, 
$\sin2\theta_d\gsim-0.6\ (-0.87)$ at one sigma (95\% CL), 
where $2\theta_d=\arg(M_{12}/M_{12}^{\rm SM})$.
This, together with constraints from $\Delta m_B$, gives the one sigma 
bounds on the CP asymmetry in semileptonic $B$ decays, $-2\times10^{-2}\lsim
a_{\rm SL}\lsim 3\times10^{-2}$.

\vskip 1 cm
\centerline{\bf Acknowledgements}
We thank Yuval Grossman and Mihir Worah for valuable comments
on the manuscript. G.B. acknowledges a post-doctoral fellowship
of the Graduiertenkolleg ``Elementarteilchenphysik bei 
mittleren und hohen Energien" of the University of Mainz.
Y.N. is supported in part by the United States $-$ Israel Binational
Science Foundation (BSF) and by the Minerva Foundation (Munich).

\listrefs
\end